\documentstyle [12pt] {article}

\catcode`@=12
\begin{document}
\thispagestyle{empty}
\begin{flushright} UCRHEP-T140\\March 1995\
\end{flushright}
\vspace{0.5in}
\begin{center}
{\Large \bf Possible Revelation of Seesaw Mass Pattern\\
in Solar and Atmospheric Neutrino Data\\}
\vspace{1.5in}
{\bf Ernest Ma\\}
{\sl Department of Physics, University of California, Riverside,
California 92521\\}
\vspace{0.1in}
{\bf J. Pantaleone\\}
{\sl Department of Physics, University of Alaska, Anchorage, Alaska 99508}
\vspace{1.5in}
\end{center}
\begin{abstract}\
Assuming the solar and atmospheric neutrino deficits to be due to neutrino
oscillations, it is shown that the $3 \times 3$ mass matrix spanning
$\nu_e, \nu_\mu$, and $\nu_\tau$ may have already revealed a seesaw mass
pattern.  Also, this matrix is the natural reduction of a simple $5 \times 5$
seesaw mass matrix with one large scale, the $4 \times 4$ reduction of
which predicts that a fourth neutrino would mix with $\nu_e$ and
$\nu_\mu$ in such a way that $\nu_\mu \rightarrow \nu_e$ oscillations may
occur just within the detection capability of the
LSND (Liquid Scintillator Neutrino Detector) experiment.
\end{abstract}
\newpage
\baselineskip 24pt

Whether or not neutrinos have mass is clearly a fundamental issue in
particle physics and astrophysics.  There is now a good deal of indirect
evidence from measurements of the solar $\nu_e$ flux\cite{1,2,3,4} and
the atmospheric $\nu_\mu/\nu_e$ ratio\cite{5,6} that neutrinos oscillate
from one to another.  This means that $\nu_e$ and $\nu_\mu$ are not mass
eigenstates; each is rather an admixture of two or more mass eigenstates.
Under the simplifying assumption that only two neutrinos are
involved in each case, probable central values for the mass-squared
difference and the mixing angle are
\begin{equation}
\Delta m^2 \simeq 4 \times 10^{-6}~{\rm eV^2}, ~~~ \sin^2 2 \theta \simeq
0.01,
\end{equation}
from analyzing\cite{7} solar data, assuming the small-angle,
nonadiabatic solution, and
\begin{equation}
\Delta m^2 \simeq 10^{-2}~{\rm eV^2}, ~~~ \sin^2 2 \theta \simeq 0.5,
\end{equation}
from analyzing\cite{8} atmospheric data.

Given only the above information, it is clearly not possible to reconstruct
unambiguously the underlying $3 \times 3$ neutrino mass matrix.  However,
if a certain empirical relationship can be found among the four parameters
listed above, an important insight may be gained as to the form of this
mass matrix.  Such an example is already very well-known in the case of
the quark mass matrix.  It was pointed out many years ago\cite{9} that
the empirical relationship
\begin{equation}
\sin^2 \theta_C \simeq m_d/m_s
\end{equation}
may be obtained with a $2 \times 2$ mass matrix of the form
\begin{equation}
{\cal M} = \left[ \begin{array} {c@{\quad}c} 0 & a \\ a & b \end{array}
\right].
\end{equation}
This simple observation has generated over the years an enormous literature
on quark mass matrices.  It is an especially active field of research in
the past two or three years.  In the jargon of neutrino physics, the form
of $\cal M$ in Eq.~(4) is called seesaw\cite{10}.

We propose here that Eq.~(4) is also applicable to the $2 \times 2$ matrix
spanning $\nu_\tau$ and the linear combination $\nu_\mu \cos \theta +
\nu_e \sin \theta$, where $\theta$ is the angle measured in atmospheric
neutrino oscillations.  The orthogonal linear combination $\nu_e \cos \theta
- \nu_\mu \sin \theta$ is assumed to be massless.  The $3 \times 3$ neutrino
mass matrix spanning $\nu_e, \nu_\mu, \nu_\tau$ is then given by
\begin{equation}
{\cal M}_\nu = \left[ \begin{array} {c@{\quad}c@{\quad}c} b \sin^2 \theta &
b \sin \theta \cos \theta & a \sin \theta \\ b \sin \theta \cos \theta &
b \cos^2 \theta & a \cos \theta \\ a \sin \theta & a \cos \theta & 0
\end{array} \right].
\end{equation}
The eigenvalues of ${\cal M}_\nu$ are 0, $-a^2/b$, and $b$ for $a << b$.
Let the corresponding mass eigenstates be $\nu_1, \nu_2$, and $\nu_3$. Then
the mixing matrix $U$ is easily obtained:
\begin{equation}
\left( \begin{array} {c} \nu_e \\ \nu_\mu \\ \nu_\tau \end{array} \right)
= \left( \begin{array} {c@{\quad}c@{\quad}c} \cos \theta & -(a/b) \sin \theta
& \sin \theta \\ -\sin \theta & -(a/b) \cos \theta & \cos \theta \\ 0 & 1 &
a/b \end{array} \right) \left( \begin{array} {c} \nu_1 \\ \nu_2 \\ \nu_3
\end{array} \right).
\end{equation}
Following the results of the two-neutrino analyses given in Eqs. (1) and (2),
let $b = 0.1$ eV and $a = 0.014$ eV, then $a^2/b \simeq 2 \times 10^{-3}$
eV, hence $\Delta m_{12}^2 \simeq 4 \times 10^{-6}~{\rm eV}^2$ and
$\Delta m_{13}^2 = 10^{-2}~{\rm eV}^2$.
Now choose $\sin^2 2 \theta = 0.5$ as given in Eq.~(2), then the
effective $\sin^2 2 \theta_{12} \equiv 4 |U_{e2}|^2 |U_{e1}|^2 /
[|U_{e1}|^2 + |U_{e2}|^2 ]^2 $
for solar neutrino oscillations is \underline {predicted} to be
\begin{equation}
\sin^2 2 \theta_{12} \simeq (a/b)^2 \sin^2 2 \theta / \cos^4 \theta
\simeq 0.014,
\end{equation}
which roughly matches the value given in Eq.~(1).  This marks the first time
that a possible seesaw mass pattern in neutrino physics has been identified.

Since $\Delta m^2_{12} << \Delta m^2_{13}$ and $a/b << 1$, the atmospheric
neutrino deficit\cite{5,6} is explained by a simple
two-neutrino oscillation between $\nu_\mu$ and $\nu_e$.
However the solar neutrino deficit comes from the
oscillations of all three neutrinos.
If matter effects \cite{11} could be
neglected, the disappearance probability of solar neutrinos would be
given by
\begin{equation}
1 - P(\nu_e \rightarrow \nu_e) = 2 |U_{e3}|^2 |U_{e1}|^2 +
2 |U_{e3}|^2 |U_{e2}|^2 +
2 |U_{e2}|^2 |U_{e1}|^2 \left( 1  -
\cos {{t \Delta m^2_{12}} \over {2p}} \right),
\end{equation}
where $U_{\alpha i}$ are the elements of the mixing matrix in Eq. (6).
We see that the effects of the heaviest mass parameter average out,
but a dependence on $|U_{e3}|^2$ remains.  Only in the limit of
vanishing $|U_{e3}|^2 = \sin^2 \theta$ is the two-neutrino approximation
valid for solar neutrinos.  This general observation is also true when
matter effects are included\cite{12}.
Since a large $\sin^2 \theta$ is required by the atmospheric data,
the naive comparison of Eq.~(7) with Eq.~(1) will have to be modified.

A three-neutrino analysis of all the solar neutrino data\cite{1,2,3,4}
is shown in Figs. (1a), (1b), and (1c).  The method is the same as that of an
earlier analysis\cite{13} where a continuous range of $|U_{e3}|^2$ values is
considered.  Three values of $\Delta m_{13}^2$
and $\sin^2 2 \theta$ which fit the atmospheric neutrino data have been chosen,
and their predictions for $\Delta m_{12}^2$ and $\sin^2 2 \theta_{e2}
\equiv 4 |U_{e2}|^2 (1-|U_{e2}|^2)$
(dashed line) have been compared to the solar data constraints (solid
contours).
For Fig.~(1a), a small value of $\sin^2 2 \theta$ is chosen.  There the
allowed solar neutrino solutions are close to those of the two-neutrino
analysis, and the predictions of Eq.~(5) match the small-angle, nonadiabatic
solution.
For Figs.~(1b) and (1c), two larger values of $\sin^2 2 \theta$ are chosen
and the allowed solar neutrino solutions are very different from those
found in the two-neutrino analysis.  Here the predictions of Eq. (5) match an
adiabatic solution which is allowed only in the three-neutrino analysis.
As shown in Fig. (2), Eqs. (5) and (6) are satisfied in different ways
across the parameter region indicated by the atmospheric data.
Although we do not include the region allowed by the new Kamiokande
multi-GeV data\cite{6} which is in conflict with Frejus \cite{14}, solutions
exist there as well.

Consider now the possible origin of the $3 \times 3$ neutrino mass matrix
${\cal M}_\nu$ of Eq.~(5).  The most natural way is to find it as the
seesaw\cite{10} reduction of a larger matrix which includes a higher
mass scale.  A very simple solution is the following
$5 \times 5$ mass matrix
\begin{equation}
{\cal M}_5 = \left[ \begin{array} {c@{\quad}c@{\quad}c@{\quad}c@{\quad}c}
0 & 0 & 0 & m_1 & 0 \\ 0 & 0 & 0 & m_2 & 0 \\ 0 & 0 & 0 & 0 & m_3 \\
m_1 & m_2 & 0 & 0 & m_4 \\ 0 & 0 & m_3 & m_4 & m_5 \end{array} \right]
\end{equation}
spanning $\nu_e, \nu_\mu, \nu_\tau, \nu_S$, and $N$, where the last two
are singlet Majorana neutrinos.  In the limit of very large $m_5$,
the seesaw reduction of ${\cal M}_5$ is
\begin{equation}
{\cal M}_4 = \left[ \begin{array} {c@{\quad}c@{\quad}c@{\quad}c}
0 & 0 & 0 & m_1 \\ 0 & 0 & 0 & m_2 \\ 0 & 0 & -m_3^2/m_5 & -m_3m_4/m_5 \\
m_1 & m_2 & -m_3m_4/m_5 & -m_4^2/m_5 \end{array} \right].
\end{equation}
Assume now that $m^2_4/m_5$ is the dominant term in ${\cal M}_4$, then a
second seesaw reduction gives the following $3 \times 3$ matrix:
\begin{equation}
{\cal M}_3 = \left[ \begin{array} {c@{\quad}c@{\quad}c} m_1^2m_5/m_4^2 &
m_1m_2m_5/m_4^2 & -m_1m_3/m_4 \\ m_1m_2m_5/m_4^2 & m_2^2m_5/m_4^2 &
-m_2m_3/m_4 \\ -m_1m_3/m_4 & -m_2m_3/m_4 & 0 \end{array} \right],
\end{equation}
which matches exactly ${\cal M}_\nu$ of Eq.~(5) with the identification
\begin{equation}
b = {{(m_1^2+m_2^2)m_5} \over m_4^2}, ~~~ \tan \theta = {m_1 \over m_2},
{}~~~ a = - {{m_3 \sqrt {m_1^2+m_2^2}} \over m_4}.
\end{equation}
With the further requirement that $m_3m_4 << m_5 \sqrt {m_1^2+m_2^2}$ so
that $a << b$, we then obtain Eq.~(6).  Since ${\cal M}_3 (={\cal M}_\nu)$
reduces further to $\cal M$ of Eq.~(4), one might even speculate that the
present neutrino data are indicative of \underline {three}
successive seesaw reductions.

If we take ${\cal M}_5$ (and thus ${\cal M}_4)$ seriously, then we have one
more very interesting prediction.  There should be a fourth neutrino ($\nu_S$)
with mass given by $-m_4^2/m_5$ and mixing to $\nu_e$ and $\nu_\mu$
given by $-m_1m_5/m_4^2$ and $-m_2m_5/m_4^2$ respectively.  Since
\begin{equation}
\left( {m_4^2 \over m_5} \right) \left( {{m_1m_5} \over m_4^2} \right)
\left( {{m_2m_5} \over m_4^2} \right) = \left( {{(m_1^2+m_2^2)m_5} \over m_4^2}
\right) \left( {m_1 \over \sqrt {m_1^2+m_2^2}} \right) \left( {m_2 \over
\sqrt {m_1^2+m_2^2}} \right),
\end{equation}
we predict that there should be additional oscillations between $\nu_e$ and
$\nu_\mu$ through this fourth neutrino mass eigenstate with restricted
values of $\Delta m^2$ and $\sin^2 2 \theta$ such that their product is the
same as that for atmospheric oscillations.  Using the numerical results
obtained earlier in this paper, we would then predict
\begin{equation}
3 \times 10^{-3}~{\rm eV}^2 < \Delta m^2 \sin^2 2 \theta < 10^{-2}~{\rm eV}^2.
\end{equation}
Part of this region is excluded by the E776 neutrino
experiment at Brookhaven National Laboratory\cite{16}, but
another part ($\Delta m^2 \sim 1~{\rm eV}^2, \sin^2 2 \theta \sim 5 \times
10^{-3}$) lies within the detection capability of the LSND (Liquid
Scintillator Neutrino Detector) experiment at Los Alamos National Laboratory.
We also note that the upper bound in the above comes from reactor
experiments\cite{15}.  Relaxing it would allow a larger $\Delta m^2 ( \sim
6~{\rm eV}^2)$ which is preferred by the preliminary LSND results,
but in conflict with the published E776 data.

To obtain ${\cal M}_5$, the following softly broken discrete Z$_3$
symmetry may be considered.  Let $\nu_e, \nu_\mu, \nu_\tau, \nu_S$, and
$N$ transform as $\omega, \omega, 1, \omega^2$, and 1 respectively,
where $\omega^3 = 1$.  Then $m_1, m_2$, and $m_3$ come from the vacuum
expectation value of the standard Higgs doublet, and $m_5$ is allowed by
Z$_3$.  However, $m_4$ breaks Z$_3$ softly and so does the diagonal mass
term for $\nu_S$, but as long as the latter is much smaller than $m_4^2/m_5$,
the reduction to ${\cal M}_\nu$ proceeds as before.  With four light
neutrinos, the nucleosynthesis bound\cite{17} of $N_\nu < 3.3$ is an important
constraint.  Although $\nu_S$ is a singlet neutrino, it mixes with
other neutrinos and may contribute significantly to $N_\nu$ through
oscillations\cite{18}. In the context of ${\cal M}_4$, this constraint
implies that $\sqrt {m_1^2 + m_2^2}$ has to be very much smaller than
$m_4^2/m_5$, hence the $\sin^2 2 \theta$ parameter in Eq.~(14) is too small
to be in the range of the LSND detection capability.  In other words,
the simplest extension of our basic proposal to four neutrinos cannot
accommodate simultaneously the LSND results and nucleosynthesis.
The latter constraint can be avoided if we switch $\nu_\tau$ and $\nu_S$
in ${\cal M}_5$, but then the underlying theory becomes much more involved.
For example, the $\nu_e \nu_\tau$ and $\nu_\mu \nu_\tau$ mass terms must
now come from either a Higgs triplet or a radiative mechanism as recently
proposed\cite{19}.  Details will be given elsewhere.

In conclusion, we have shown in the above how present data on solar and
atmospheric neutrinos may have revealed a seesaw mass pattern for the three
known neutrinos.  This seesaw scenario requires the solar neutrino deficit
to involve the oscillations of all three neutrinos.  Hence a much more
extensive numerical analysis is needed\cite{13} beyond that of the usual
two-neutrino assumption\cite{7}.  It is interesting to note that with only
two neutrinos, the adiabatic solution is not allowed by the combined solar
data, but with three neutrinos, it becomes allowed if the mixing of the
third neutrino with $\nu_e$ is large enough.

We have also shown that a simple $5 \times 5$ mass matrix with one large
scale reduces naturally to the desired $3 \times 3$ ${\cal M}_\nu$ of Eq.~(5).
Furthermore, the intermediate $4 \times 4$ reduction predicts a fourth
neutrino which mixes with $\nu_e$ and $\nu_\mu$ in such a way that $\nu_\mu
\rightarrow \nu_e$ oscillations may occur just within the detection
capability of the LSND experiment.
\vspace{0.5in}

\begin{center}{ACKNOWLEDGEMENT\\}
\end{center}

This work was supported in part by the U.S. Department of Energy under Grant
No. DE-FG03-94ER40837.

\newpage
\bibliographystyle{unsrt}

\begin{thebibliography}{99}
\bibitem{1}  R. Davis, Jr. {\it et al}., Ann. Rev. Nucl. and Part. Sci.
{\bf 39}, 467 (1989).
\bibitem{2}  K.S. Hirata {\it et al}., Phys. Rev. Lett. {\bf 63}, 16 (1989);
{\bf 65}, 1297 (1990); {\bf 66}, 9 (1991).
\bibitem{3}  A.I. Abazov {\it et al}., Phys. Rev. Lett. {\bf 67}, 3332 (1991).
\bibitem{4}  P. Anselmann {\it et al}., Phys. Lett. {\bf B314}, 445 (1993);
{\bf B327}, 377 (1994).
\bibitem{5}  R. Becker-Szendy {\it et al}., Phys. Rev. D {\bf 46}, 3720 (1992).
\bibitem{6}  K.S. Hirata {\it et al}., Phys. Lett. {\bf B280}, 146 (1992);
Y. Fukuda {\it et al., ibid.} {\bf B335}, 237 (1994).
\bibitem{7}  See for example S.A. Bludman {\it et al}., Phys. Rev. D {\bf 47},
2220 (1993).
\bibitem{8}  See for example W. Frati {\it et al}., Phys. Rev. D {\bf 48},
1140 (1993).
\bibitem{9}  S. Weinberg, Ann. N.Y. Acad. Sci. {\bf 38}, 185 (1977); see
also E. Ma, Phys. Rev. D {\bf 43}, R2761 (1991).
\bibitem{10}  M. Gell-Mann, P. Ramond, and R. Slansky, in {\it Supergravity},
Proceedings of the Workshop, Stony Brook, New York, 1979, edited by P.
van Nieuwenhuizen and D.Z. Freedman (North-Holland, Amsterdam, 1979), p. 315;
T. Yanagida, in {\it Proceedings of the Workshop on the Unified Theory and
the Baryon Number in the Universe}, Tsukuba, Ibaraki, Japan, 1979, edited by
O. Sawada and A. Sugamoto (KEK Report No. 79-18, Tsukuba, Japan, 1979).
\bibitem{11}  S.P. Mikheyev and A.Yu. Smirnov, Yad. Fiz. {\bf 42}, 1441 (1985)
[Sov. J. Nucl. Phys. {\bf 42}, 913 (1985)]; Nuovo Cimento {\bf 9C}, 17 (1986);
L. Wolfenstein, Phys. Rev. D {\bf 17}, 2369 (1978).
\bibitem{12}  T.K. Kuo and J. Pantaleone, Rev.
Mod. Phys. {\bf 61}, 937 (1989).
\bibitem{13} D. Harley, T.K. Kuo and J. Pantaleone, Phys. Rev. D {\bf 47},
4059 (1993).
\bibitem{14} Ch. Berger {\it et al.}, Phys. Lett. {\bf B227}, 489 (1989);
{\bf B245}, 305 (1990).
\bibitem{15} G.S. Vidyakin {\it et al}., Sov. Phys.
JETP {\bf 71} 424 (1990); G. Zacek {\it et al}., Phys. Rev. D {\bf 34},
2621 (1986).
\bibitem{16} L. Borodovsky {\it et al.}, Phys. Rev. Lett. {\bf 68}, 274 (1992).
\bibitem{17} See for example T. Walker {\it et al.}, Astrophys. J. {\bf 376},
51 (1991).
\bibitem{18} See for example X. Shi {\it et al.}, Phys. Rev. D {\bf 48},
2563 (1993).
\bibitem{19} E. Ma, Phys. Rev. D {\bf 51} (Rapid Communication), in press.
\end{thebibliography}

\raggedbottom
\pagebreak

\begin{center}
Figure Captions
\end{center}

\begin{description}

\item[Figs. (1)] Plots of $\Delta m_{12}^2$ versus $\sin^2 2
\theta_{e2} \equiv 4 |U_{e2}|^2 (1-|U_{e2}|^2)$
for various values of $\sin^2 2 \theta$.
The solid contours surround the parameter region allowed
by the solar neutrino data \cite{1,2,3,4} at 90\% confidence level.
The dashed line gives the prediction
of Eq.~(5) for a specific value of $\Delta m_{13}^2$
which explains the atmospheric neutrino deficit.

\item[~~~(a)] $\sin^2 2 \theta = 0.35$
and $\Delta m_{13}^2 = 3.0 \times 10^{-2}$ eV$^2$.

\item[~~~(b)] $\sin^2 2 \theta = 0.50$ and
$\Delta m_{13}^2 = 1.0 \times 10^{-2}$ eV$^2$.

\item[~~~(c)] $\sin^2 2 \theta = 0.75$
and $\Delta m_{13}^2 = 4.0 \times 10^{-3}$ eV$^2$.

\item[Fig. (2)]  Plot of $\Delta m_{13}^2$ versus $\sin^2 2 \theta$.
The dashed contours surround the parameter region allowed
by sub-GeV atmospheric neutrino measurements\cite{6} but excluded
by the Frejus data\cite{14} and
below constraints from reactor measurements\cite{15}.
The shaded regions are where Eq.~(5) can satisfy the
solar neutrino data; the left shaded region corresponds to the
small-angle nonadiabatic solution and the right shaded region
corresponds to the adiabatic solution.  In between these two
regions is a small area where it is not possible to satisfy
the solar neutrino data.

\end{description}

\end{document}